# The Luneburg–Lissajous lens

HUIYAN PENG [1,2], HUASHUO HAN [1,2], PINCHAO HE [1,2], KEQIN XIA [1,2], JIAXIANG ZHANG [1,2], XIAOCHAO LI [1,2], QIAOLIANG BAO [3], YING CHEN [1,2, †], and HUANYANG CHEN [1,2, ‡]

[1] *Institute of Electromagnetics and Acoustics and Key Laboratory of Electromagnetic Wave Science and Detection Technology, Xiamen University Xiamen 361005, China*

[2] *Department of Electrical and Electronics Engineering, Xiamen University Malaysia, 43900 Sepang, Selangor, Malaysia*

[3] *Department of Materials Science and Engineering, ARC Centre of Excellence in Future Low-Energy Electronics Technologies (FLEET), Monash University, Clayton, Victoria 3800, Australia*

PACS `42.15.Dp`
PACS `42.15.Eq`

**Abstract** – We design a new absolute optical instrument by composing Luneburg lens and Lissajous lens, and analyze its imaging mechanism from the perspective of simple harmonic oscillations. The imaging positions are determined by the periods of motions in x and y directions. Besides, instruments composed with multi parts are also devised, which can form imaging or self-imaging as long as the motion periods of x and y directions are satisfied to similar conditions. Our work provides a new way to analyze the imaging of different lens by simply dissociating the equations of motions, and reveal the internal mechanism of some absolute optical instruments.

**Introduction.** - Absolute optical instruments are important lenses with perfect imaging or self-imaging in geometric optics [1, 2, 3]. It is very important in conformal cloaking designs [3, 4, 5]. The famous lenses are Maxwell's fish-eye lens [6], Etaon lens [7] and Luneburg lens [8]. Most of them are of rotation symmetry. Recently, Lissajous lens [9] has also been studied to be such kind of lenses, yet without rotation symmetry. Conformal transformation could also be performed on Mikaelian lens [10] to obtain lenses without rotation symmetry [11]. A general lens could be achieved by the analogy of geometric optics and classical mechanics based on Hamilton-Jacobi equation [12]. In fact, the Luneburg lens and the Lissajous lens share the same motions of equations, which are simply the harmony oscillations [2]. Hence, if we could separately construct different harmony oscillations in x and y directions, by matching well each of their periods, new kind of absolute optical instruments could be obtained.

In the paper, we simply study the compositions of Luneburg lens and/or Lissajous lens with matching boundaries, and study their imaging properties based on harmony oscillations. We give a very simple picture for this kind of complicated lenses.

**Results.** - Let's start from a well-known Luneburg lens expressed as n(x,y)=$\sqrt{2-x^2-y^2}$, and the trajectory is shown in FIG.1 (a), where a closed elliptic light path can be seen inside the Luneburg lens. Similarly, for the Lissajous lens [9] n(x,y)= $\sqrt{2-x^2/a^2-y^2/c^2}$ with a=1, c=2, as displayed in FIG. 1(b), the index at the boundary of this elliptical lens (the yellow ellipse with semiaxes √2 and √8) is zero, the trajectory is also a closed one. As it is known, the above two classical lenses are absolute optical instruments, namely, rays emitting from any point $(x_0, y_0)$ in space will later converge to another point and form imaging (or self-imaging). We next design a Luneburg-Lissajous (L-L) lens in FIG.1 (c), with the refractive index profiles n(x,y) =$\sqrt{2-x^2-y^2}$ in the upper space (y>0) and n(x,y) = $\sqrt{2-x^2/a^2-y^2/c^2}$ in the lower space (y<0). The continual refractive index condition at the interface of y=0 should be satisfied to keep light without either refraction or reflection at the interface. FIG. 1(c) shows an example of such a L-L lens with a=1 and c=2. Interestingly, a closed path can still be found in such a composite lens. Light emerging from $(x_0, y_0)$ point will later return back to the original point after going through several periods. The closed path indicates that this L-L lens may be an effective absolute optical instrument.

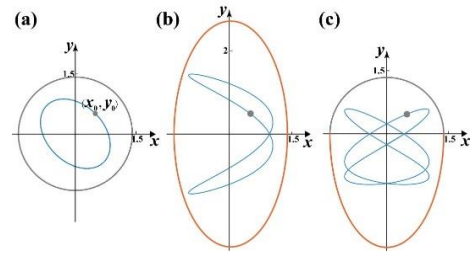

Fig. 1 : Ray trajectories in (a) the Luneburg lens, (b)

† E-mail: `irusying1218@stu.xmu.edu.cn`
‡ E-mail: `kenyon@xmu.edu.cn`

the Lissajous lens with a=1, c=2, (c) the Luneburg-Lissajous lens with a=1, c=2. The outer solid lines denote the *n*=0 boundaries, with different colors represent different lenses. The grey points denote the source at the coordinate of $(x_0, y_0)$ inside the lens.

In FIG. 2, we first analyze the imaging mechanisms of Lissajous lens from the perspective of simple harmonic oscillations (SHO), based on the analogy of light to classical mechanics [2]. The emission point $(x_0, y_0)$ is set at (0.5, 0.5) in the first quadrant for example, as shown by the grey points. For the refractive index distribution of a two-dimensional (2D) Lissajous lens $n(x,y)= \sqrt{2 - x^2/a^2 - y^2/c^2}$, the solved motion curves of SHO along x and y directions can be written as

$$\begin{cases} x = x_0 \cos(t/a) + A \sin(t/a) \\ y = y_0 \cos(t/c) + B \sin(t/c) \end{cases}$$

(1)

where t is the time parameter from classical mechanics, A and B are constants depending on the initial emission angle $\theta$ (i.e., arctan(B/A) as the tangential angle $\theta$), and $(x_0, y_0)$ is the coordinate of the source. We can see that the motions in x and y directions have the period of $2a\pi$ and $2c\pi$, respectively. The full period is then equals to $2N\pi$, where N is the least common multiple of a and c. When a=1, c=2, the ray trajectories in the lens can be seen in FIG.2 (a), where three rays are emitted from the original point and imaged at $(x_0, -y_0)$, following mirror symmetry along x-axis. We also plot two other cases with a=1, c=3 and a=2, c=1, and the optical paths are displayed in FIG. 2(c) and (e). Interestingly, different imaging positions are obtained. For the Lissajous lens with a=1, c=3, the centrosymmetric imaging is achieved at $(-x_0, -y_0)$, while for that with a=2, c=1, imaging of mirror symmetry along y-axis is shown. Therefore, three kinds of imaging position are observed in Lissajous lens. In order to explore the relationships between the imaging mechanism and parameters a and c, we analytically plot the curves of SHO along x and y directions based on Eq. (1), as shown in FIG. 2(b), (d) and (f). The solid curves indicate the motions in x direction, while the dashed curves represent those in y direction. Firstly, for the SHO of Lissajous lens with a=1, c=2 (see FIG. 2(b)), when emitting from the grey point $(x_0, y_0)$ at t = 0, the motions in x direction have a period of $2\pi$ for the three paths, and that in y direction is $4\pi$. We find that these intersections locating at x and y curves are exactly the coordinate points $(x_0, -y_0)$ and $(x_0, y_0)$, corresponding to the results in FIG. 2(a). The imaging time parameter is at $2\pi$, while the full period is $4\pi$. When it comes to the case of a=1 and c=3 in Eq. (1), the oscillation periods in x and y direction are $2\pi$ and $6\pi$ respectively, as shown in FIG. 2(d). The imaging point appears at the time of $3\pi$, i.e., $(-x_0, -y_0)$ and the full period is $6\pi$, coming back to $(x_0, y_0)$. Similarly, the light oscillations for the case of a=2 and c=1 are shown in FIG. 2(f), with the imaging time of $2\pi$, i.e., at the black dots related to $(-x_0, y_0)$. We can see that the analysis of SHO gives a nice explanation of the imaging mechanism, which is determined by the periodicities of SHO in both x and y directions. The position of imaging time is at $t = N\pi$. Again, N is least common multiple of a and c.

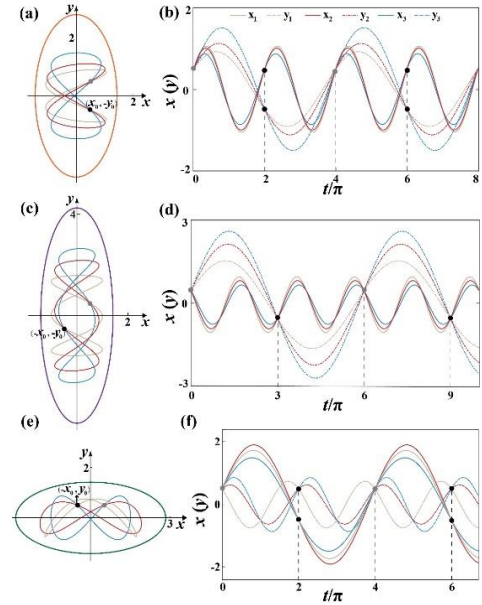

Fig. 2 : Imaging at different positions for the Lissajous lens with (a, b) a=1, c=2, (c, d) a=1, c=3, (e, f) a=2, c=1. The grey dots denote the source position $(x_0, y_0)$ in the first quadrant, and the black dots represent the imaging points. Three light paths denoting $x_1$, $y_1$; $x_2$, $y_2$ and $x_3$, $y_3$ are shown in different colors.

We next come to the general L-L lens and study the ray propagation inside it. As it is shown in FIG.3, there are two parts composing it, with different refractive index profiles:

$$n_{L-L}(x, y) = \begin{cases} \sqrt{2 - x^2/a^2 - y^2/b^2} & y > 0 \\ \sqrt{2 - x^2/a^2 - y^2/c^2} & y < 0 \end{cases}$$

(2)

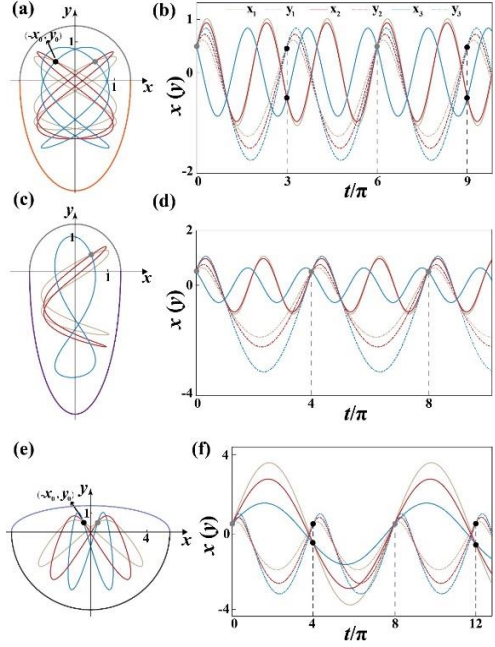

Fig. 3 : The general L-L lens with (a, b) a=1, b=1 and c=2; (c, d) a=1, b=1 and c=3; (e, f) a=4, b=1 and c=3. The left column represents the trajectories in the lens with the refractive index $n_{L-L}$, while the right column displays the oscillations along x and y directions.

The refractive index of at the boundary of y =0 is matched for every x. In terms of the SHO, the period in x direction is still 2aπ, while that in y direction has changed to (b+c) π. The full period is then equals to Mπ, where M is the least common multiple of 2a and (b+c). To explore the ray behavior inside such L-L lens and verify its imaging features, we first consider a L-L lens with a=1, b=1 and c=2 (the lens in FIG. 1(c)). The trajectories of three light paths are shown in FIG. 3(a), where light emitting from the source position ($x_0$, $y_0$) with different angles later focuses on the mirror point (-$x_0$, $y_0$) without aberration, illustrating the property of absolute optical instrument. In FIG.3 (b) the analytic harmonic motions along x and y directions are also exhibited. We clearly see that the intersections of x and y rays happen at the position of (-$x_0$, $y_0$) when t=Mπ/2=3π, and at the position of ($x_0$, $y_0$) when t=Mπ=6π, forming y-axis symmetric imaging and self-imaging. Secondly, when a=1, b=1 and c=3, the closed ray trajectories and dissociated x- and y-motion paths are shown in FIG. 3(c) and 3(d). Obviously, the perfect intersections only appear at the integer time of t=Mπ=4π, corresponding to the coordinate ($x_0$, $y_0$), which illustrates only the self-imaging phenomenon in FIG. 3(c). Note that for t=Mπ/2=2π, there is no perfect intersection, which means that there is no other image. In the last case of a=4, b=1 and c=3 in FIG. 3(e) and 3(f), the intersections are at the position of (-$x_0$, $y_0$) for t=Mπ/2=4π, and the position of ($x_0$, $y_0$) for t=Mπ=8π, i.e., the y-axis symmetric imaging and self-imaging.

Therefore, by composing two lenses together, a new absolute optical instrument is designed with various imaging positions, which are determined by the periods of SHO in x and y directions.

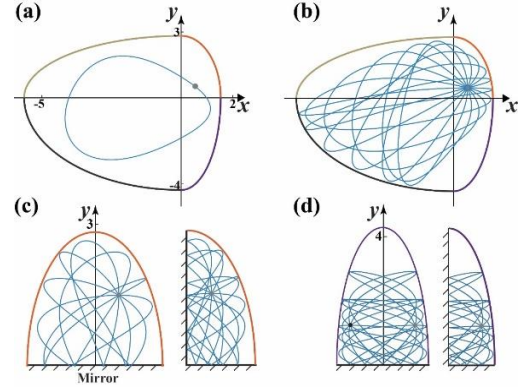

Fig. 4 : (a, b) Composed L–L lens with four different parts for (a) a closed light path; (b) a point source. (c, d) Imaging of Lissajous lens under the x-axis mirror at y =0; x- and y-axis mirrors at x=0 and y=0, based on Lissajous lenses in Fig. 2(a) and (c) respectively.

Lastly, we introduce a complicated L-L lens composed by four Lissajous lens, as denoted by four different colors in FIG. 4(a). The boundary of n = 0 is determined by the following refractive index profiles:

$$n_{L-L}(x,y) = \begin{cases} \sqrt{2 - \dfrac{x^2}{a^2} - \dfrac{y^2}{b^2}} & x>0, y>0 \\ \sqrt{2 - \dfrac{x^2}{d^2} - \dfrac{y^2}{b^2}} & x<0, y>0 \\ \sqrt{2 - \dfrac{x^2}{d^2} - \dfrac{y^2}{c^2}} & x<0, y<0 \\ \sqrt{2 - \dfrac{x^2}{a^2} - \dfrac{y^2}{c^2}} & x>0, y<0 \end{cases}$$

(3)

where the continual refractive index conditions are satisfied at both x =0 and y = 0 for the above similar reason. Here we take a=1, b=2, c=3, d=4 for example, the L-L lens is shown in FIG. 4(a). We can see that the light can still return to the emission point and forms a closed trajectory. For such a special lens, the self-imaging character is displayed in FIG. 4(b), which can be comprehended easily through the period of SHO above. For the oscillation in x direction, the period is (a+d) π, while for the y direction, the period is (b+c)π. The self-imaging happens at t=Mπ, where M is the least common multiple of (a+d) and (b+c). For this case, M=5. In addition, we can also construct other lenses using reflection mirrors. For example in FIG. 4(c) and 4(d), two cases with reflection mirrors are discussed to further study the imaging properties based on the Lissajous lenses in FIG. 2(a) and 2(c). As shown in FIG. 4(c), when a x-axis mirror (y = 0) is added, the previous imaging at ($x_0$, -$y_0$) is mapped to the emission point and only self-imaging is observed. Meanwhile, when two mirrors at both x = 0 and y = 0 are added, the rays are reflected back with self-imaging as well. However, for the case in FIG.2 (c), the x-axis mirror at y = 0 can map the imaging at the point of (-$x_0$, -$y_0$) to the point (-$x_0$, $y_0$), as denoted by the black dot of FIG. 4(d). However, for mirrors at both x = 0 and y = 0, there is only self-imaging effect.

**Conclusion.** - In this work, we propose a composed Luneburg–Lissajous (L-L) lens with perfect imaging, i.e., an absolute instrument. The imaging mechanisms of Lissajous lens and L-L lens are studied by harmonic oscillations in x and y directions. The self-imaging happens at the parameter time of the least common multiple time of the periods of x and y directions. The imaging happens at half of the self-imaging time, which have to be verified from the oscillations in x and y directions. A complicated L-L lens and the effect of different reflection mirrors on imaging are further discussed. This method can be easily extended to three dimensions. Our study therefore enriches the family of absolute optical instruments, especially for Luneburg lens and Lissajous lens. They could be realized by either from curve surfaces, such as geodesic lenses [13, 14], or gradient index lenses with required dielectric profiles [15, 16].

**Acknowledgements.** - This work was financially supported by the National Natural Science Foundation of China (Grant No. 11874311) and the Fundamental Research Funds for the Central Universities (Grant No. 20720170015). Huashuo Han, Pinchao He, Keqin Xia, and Jiaxiang Zhang are undergraduate students in Xiamen University Malaysia. We thank the support from Prof. Zhong Chen, Prof. Huiqiong Wang, Mr. Jingfeng Chen, and Miss Yuling Zheng.